\documentclass[3p,onecolumn]{elsarticle}
\pdfoutput=1
\DeclareGraphicsExtensions{.pdf}

\biboptions{sort&compress}
\usepackage[colorlinks=true,linkcolor=blue, citecolor=blue, urlcolor=blue]{hyperref}
\usepackage{natbib}
\usepackage{color}
\usepackage{multirow}
\usepackage{amssymb}
\usepackage{amsmath}
\usepackage{slashed}
\usepackage{graphicx}
\usepackage{placeins}
\usepackage{array}
\usepackage{dblfloatfix}
\usepackage{scalerel} 
\usepackage{subcaption}
\usepackage{cancel}
\usepackage{xcolor}

\newcommand\ff{$\mathrm{f}\mathrm{f}$}
\newcommand\ffbar{$\mathrm{f\bar{f}}$}
\newcommand\ffshift{\kern-.4ex\raisebox{-.43ex}{\ff}\kern-.40ex}
\newcommand\ffbarshift{\kern-.4ex\raisebox{-.43ex}{\ffbar}\kern-.40ex}

\newcommand\fifi[2]{$\mathrm{f_{#1}}\mathrm{f_{#2}}$}
\newcommand\fifibar[2]{$\mathrm{f_{#1}\bar{f}_{#2}}$}
\newcommand\fifishift[2]{\kern-.4ex\raisebox{-.43ex}{\fifi{#1}{#2}}\kern-.40ex}
\newcommand\fifibarshift[2]{\kern-.4ex\raisebox{-.43ex}{\fifibar{#1}{#2}}\kern-.40ex}

\newcommand\papa[2]{$#1#2$}
\newcommand\papabar[2]{$#1\bar{#2}$}
\newcommand\papashift[2]{\kern-.4ex\raisebox{-.43ex}{\papa{#1}{#2}}\kern-.40ex}
\newcommand\papabarshift[2]{\kern-.4ex\raisebox{-.43ex}{\papabar{#1}{#2}}\kern-.40ex}
\newcommand{\mb}{M}

\begin{document}

\begin{frontmatter}
\title{
\center{On macroscopic residual QCD force of electrodynamics}
}

\author[Charles]{Martin Spousta}
\address[Charles]{Institute of Particle and Nuclear Physics, Charles University, Prague, Czech Republic}

\begin{abstract}

We explore a connection between virtual particles of quantum electrodynamics and quantum chromodynamics (QCD) which is predicted to give rise to a residual attractive 
interaction measurable as a macroscopic force. We calculate the asymptotic behavior of relevant scattering amplitudes, perform their resummation, and analyze the sign 
of the resulting interaction. Then, we calculate the primary experimentally observable consequences of this Standard Model force. We discuss the impact of this force at 
terrestrial scales and at astrophysical scales. In particular, we quantify the impact of this force on the warm ionized medium present in galaxies and the 
intracluster medium present in cluster of galaxies.

\end{abstract}

\end{frontmatter}

\newcommand*{\PP}{\mbox{$P^\prime$}}
\newcommand*{\bfp}{\mathbf{p}}
\newcommand*{\alphares}{a} 
\newcommand*{\alphas}{\alpha_\mathrm{s}}
\newcommand*{\alphah}{\Delta\alpha_\mathrm{had}}
\newcommand*{\alphaQED}{\texttt alphaQED}
\newcommand*{\fd}{\mathrm{d}}
\newcommand*{\ra}{\rightarrow}
\renewcommand{\c}[1]{\cancel{#1}}


\newlength{\fighalfwidth}
\setlength{\fighalfwidth}{0.49\textwidth}

\section{Introduction}
\label{sec:intro}

Quantum field theory (QFT) and the principle of gauge invariance stand as fundamental pillars of the Standard Model of particle physics, successfully describing 
three out of four forces in nature. QFT explains the forces in the macroscopic world as emerging from exchanges at a microscopic level of so-called virtual 
particles. These virtual particles mediate the interaction between real particles existing at macroscopic timescales. This concept of QFT allows an extremely precise 
theoretical description of phenomena seen in nature, e.g., running coupling constants of quantum electrodynamics (QED), $\alpha$ 
\cite{Aoyama:2012wj,TOPAZ:1997ipd,KLOE-2:2016mgi}, and quantum chromodynamics (QCD), $\alphas$ \cite{FlavourLatticeAveragingGroup:2019iem,Workman:2022ynf}; anomalous 
magnetic moment of electron, $g_e$ \cite{Aoyama:2012wj,Hanneke:2008tm} and muon, $g_\mu$ \cite{Aoyama:2020ynm,Muong-2:2021ojo,Borsanyi:2020mff}; higher-order QED 
processes which have no classical analogy, such as recently measured light-by-light scattering \cite{Aaboud:2017bwk,ATLAS:2019azn}.
This paper explores a connection between QED and QCD virtual particles, which is predicted to give rise to a residual attractive force. This attractive force stems 
from gluon exchanges between strongly interacting degrees of freedom in off-shell photon.

The presence of strongly interacting degrees of freedom is well known to affect the value of $\alpha$, $g_e$ or $g_\mu$ where it is quantified by the hadronic vacuum 
polarization (HVP) of photon propagator, $\Pi^\mathrm{had}$.
Baseline calculations of HVP were done by several authors \cite{Jegerlehner:1985gq,Kniehl:1989yc,Nasrallah:1997dh}.
Since the strong interaction turns non-perturbative due to the rise of 
$\alphas$ at low energy, the low energy behavior of $\Pi^\mathrm{had}$ is 
difficult to calculate using perturbative methods. Instead, 
it can be determined via the optical theorem, which connects $\Pi^\mathrm{had}$ with the total 
cross-section for the production of hadrons in $e^+e^-$ collisions.
Besides this approach, the $\Pi^\mathrm{had}$ can be also determined from lattice QCD with increasing precision \cite{Borsanyi:2020mff,Burger:2015lqa}.

\begin{figure}[h]
\begin{center}
\includegraphics[width=0.87\textwidth]{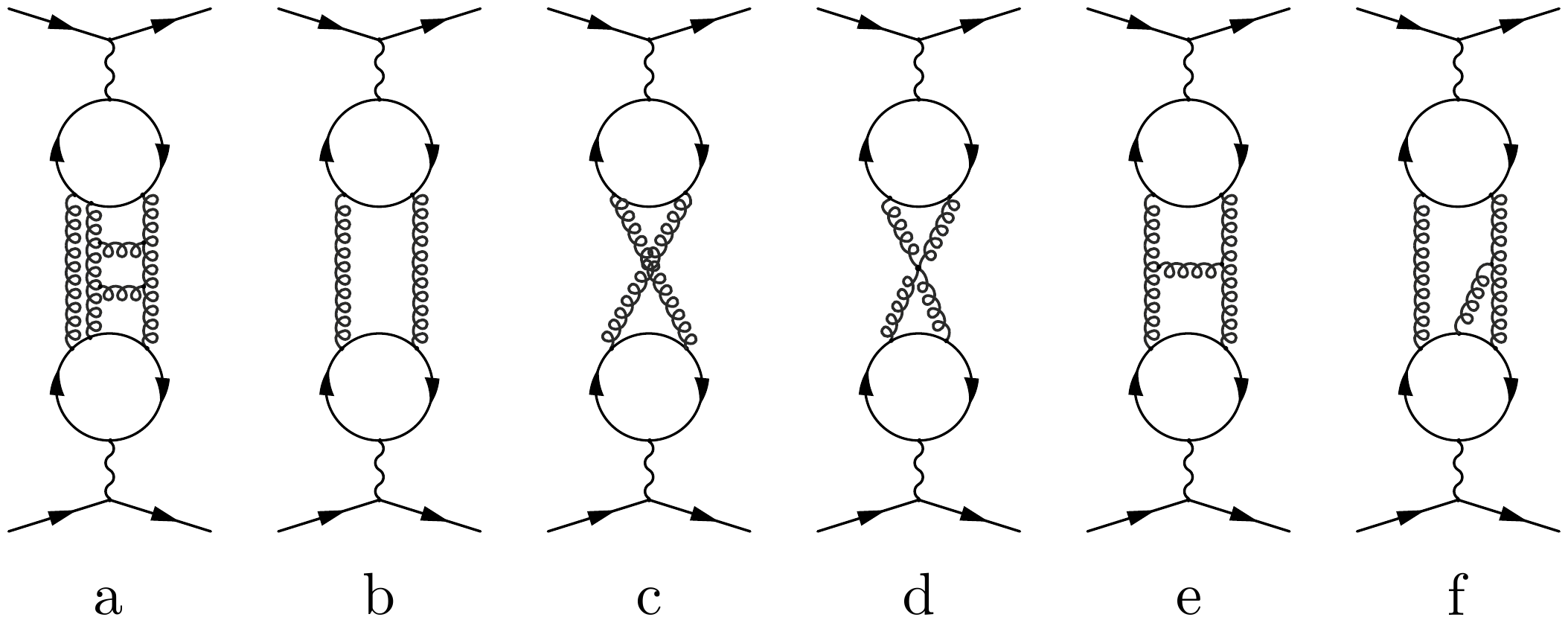}
\end{center}
\caption{
   Example of Feynman diagram for multiple gluon exchanges between two quark loops (a). Inter-quark-loop exchanges for 3-point quark loops without self-interactions 
   (b,c), with self-interactions (d,e), and exchanges between 3- and 4-point quark loops (f). Only diagrams corresponding to {\it non-abelian cases} are shown (for the 
   definition, see the text).
}
\label{fig:fig1}
\end{figure}

The contributions to $\Pi^\mathrm{had}$ at the momentum transferred ($q^2$) at the scale of hadron masses can be thought of as being due to exchanges of hadrons 
(predominantly mesons) in the loops \cite{Bauer:1977iq}. For the $q^2$ scale significantly below the hadron masses ($q^2 \ll m_\mathrm{had}$), the relevant degrees 
of freedom in HVP are quarks and gluons. These are the same degrees of freedom as those responsible for the confinement of quarks in hadrons. One may, therefore, 
expect a negligible but non-zero confining force to act between the quarks in loops inside photons surrounding charged particles as well. Since photon has no direct 
coupling to gluon, it is the quark loops which are the basic strongly interacting degrees of freedom within photon. It is multi-gluon exchanges between quarks rather 
than exchanges within quark loops which seem to be relevant for the confinement \cite{Deur:2016tte}. Thus, multi-gluon exchanges between two quark loops should represent 
the basic contribution forming this force. An example of Feynman diagram contributing to this force is shown in Figure~\ref{fig:fig1}{\color{blue}a}. This `new' force, which is, however, purely a Standard Model force, may be expected to have experimentally observable 
consequences and may obscure explanations of important observations at the scale of the gravity interaction, such as signatures of the dark matter.

In Section~\ref{sec:asy} of this paper, we calculate the asymptotic behavior of inter-quark-loop interactions. The contributions of these interactions to the 
scattering amplitude are then resummed in Section~\ref{sec:res}. The sign and strength of the interaction is discussed in Section~\ref{sec:sign}. Observable 
consequences of this residual force are then discussed in Section~\ref{sec:est}.


\section{Asymptotic behavior of inter-quark-loop interactions}
\label{sec:asy}

The quark degrees of freedom inside photons surrounding two external charges may interact via the exchange of vector particles (photons, electroweak bosons, and gluons) and 
scalar particle (Higgs). Both electroweak boson and Higgs are too heavy to appear as effective degrees of freedom in the $q^2 \ll m_\mathrm{had}^2$ region. We will, therefore, discuss 
only the exchanges of photons between two quark loops (referred to as \textit{abelian case}), and 
exchanges of gluons between two quark loops (referred to as \textit{non-abelian case}). The latter is then the case where attractive confining forces may take place.
The amplitude for these interactions can be expressed as
\begin{eqnarray}
	\label{eq:M}
	i \mathcal{M}^{\alpha\beta} &=& (i\sqrt{a})^2 \frac{-i g^{\alpha\mu}}{q^2 +i\epsilon} Y_{\mu\nu}(q) \frac{-ig^{\nu\beta}}{q^2 + i\epsilon}, \\
        \label{eq:Y}
        Y_{\mu\nu}(q) &=& \int \prod_{i} \fd u_i A_{\mu,\mu_1 \cdots \mu_{n-1}}(u_i,q) B^{\mu_1 \cdots \mu_{n-1},\nu_1 \cdots \nu_{n'-1}}(u_i, q) A'_{\nu,\nu_1 \cdots \nu_{n'-1}}(u_i,q),
\end{eqnarray}
where $A$ and $A'$ represent $n$- and $n'$-point quark loop amplitudes, respectively, $B$ represents exchanges of boson fields, and $u_i$ are unconstrained 
four-momenta (unconstrained four-momenta in quark loops are not explicitly written). We start the discussion with only the inter-loop exchanges. The intra-loop 
exchanges are mentioned later. Here, we also ignore the non-contracted wave function of incoming and outgoing charges, represented by bispinors and 
corresponding creation and annihilation operators. In the non-relativistic limit, these have an impact on the sign of the interaction, which is discussed in 
Section~\ref{sec:sign}.

For the abelian case, $n=n'$, and there are $n-2$ momenta $u_i$ connecting quark loops forming together multi-loop ($n$-loop) integral. For both abelian and 
non-abelian case the amplitude consists of at least three loops. The amplitudes $A$ and $A'$ also depend on quark masses, $m$, and $B$ depends on intermediate boson mass, 
$\mb$, which is also not explicitly written in (\ref{eq:Y}). In the abelian case, $\mb$ represents an infrared cutoff, which should be sent to 
zero at the end of the calculation. In the non-abelian case at low energies, $\mb$ is the effective gluon mass stemming from non-perturbative aspects of infrared 
finite gluon propagator and should not be neglected (see e.g. one of the latest calculations in Ref.~\cite{Horak:2022aqx}).

  One loop amplitudes may be calculated using extensions of the original Passarino-Veltman (PV) reduction \cite{Passarino:1978jh} or several new methods targeting 
mainly the LHC physics (see e.g. review \cite{Ellis:2011cr}). Calculations of multi-loop integrals remain an open problem though. Moreover, in the low-$q^2$ regime, 
amplitudes such as the one in Figure~\ref{fig:fig1}{\color{blue}a} are perturbative in $\alpha$, but non-perturbative in the non-abelian part, meaning that adding 
more gluon exchanges may bring a larger contribution to the full amplitude.
  For those reasons, we will not attempt to provide a full solution of amplitudes at a given order, but we will discuss the general behavior of these amplitudes. In 
particular, we will explore the behavior of these amplitudes in the non-relativistic limit of $q^2 \ra 0$. For some of the calculations presented in the remainder of this 
section, we used software for automated calculation of loop integrals \cite{Patel:2016fam} interfaced with Python scripts allowing to sort or isolate various 
contributions in an efficient way.

The first possible contribution with $n=n'=2$ and single gluon exchange between quark loops not captured by the formula (\ref{eq:Y}) is forbidden due to color-charge 
conservation. 
The first contribution we will analyze is therefore the contribution with $n=n'=3$, that is the exchange between two 3-point quark loops, which appears both in 
the abelian and non-abelian case. There are two diagrams contributing, Figures~\ref{fig:fig1}{\color{blue}b} and \ref{fig:fig1}{\color{blue}c}, forming three-loop 
amplitudes $Y_{\mu\nu}^b$ and $Y_{\mu\nu}^c$, respectively. The space-time part of amplitude $Y_{\mu\nu}^b$ is given by
  \begin{eqnarray}
  \nonumber
  Y_{\mu\nu}^b = \alpha_{(\mathrm{s})}^2 \int \fd^4 u_1 \mkern-35mu
  & & \int \fd^4 p \mkern+5mu \mathrm{Tr} \bigg\{ \gamma_\mu \frac{i}{\c{p}+\c{q}-m} \gamma_{\mu_1} \frac{i}{\c{p}+\c{u}_1+\c{q}-m} \gamma_{\mu_2} \frac{i}{\c{p}      -m} \bigg\} 
      \frac{-i g_{\mu_1 \nu_1}}{(u_1+q)^2-\mb^2}   \frac{-i g_{\mu_2 \nu_2}}{u_1^2-\mb^2} \\
  & & \int \fd^4 p' \mkern+5mu \mathrm{Tr}\bigg\{\gamma_\nu \frac{i}{\c{p'}      -m} \gamma_{\nu_2} \frac{i}{\c{p'}+\c{u}_1+\c{q}-m} \gamma_{\nu_1} \frac{i}{\c{p'}+\c{q}-m} \bigg\},
  \end{eqnarray}
  where $\alpha_{(\mathrm{s})}$ is $\alpha_{\mathrm{s}}$ and $\alpha$ for abelian and non-abelian case, respectively.
  The amplitudes $Y_{\mu\nu}^b$ and $Y_{\mu\nu}^c$ come out zero at various levels. First, they are zero due to Furry's theorem (the odd number of bosons attached to 
the quark loop) for the abelian case as well as for the non-abelian case where the color structure does not protect the amplitude (trace over two Gell-Mann 
matrices is symmetrical under the exchange of color indices).
  Aside from that, the individual amplitudes may be calculated exactly in the limit $q^2 \ra 0$. When summed up, the two contributions come out with equal magnitudes 
but opposite signs, and they sum up to zero even at this level. Finally, each amplitude is equal to zero separately for $\mb^2 \ra 0$. 
For the non-abelian case, there 
are two more amplitudes with $n=n'=3$, namely Figure~\ref{fig:fig1}{\color{blue}d}, labeled $Y_{\mu\nu}^d$, and Figure~\ref{fig:fig1}{\color{blue}e}, labeled $Y_{\mu\nu}^e$ (and amplitude with crossings). 
Again, these amplitudes come out zero due to Furry's theorem. Furry's theorem renders zero also amplitudes with $n=3$ and $n'>3$, such as $Y_{\mu\nu}^f$ shown in Figure~\ref{fig:fig1}{\color{blue}f}.

\begin{figure}[h]
\begin{center}
\includegraphics[width=0.6\textwidth]{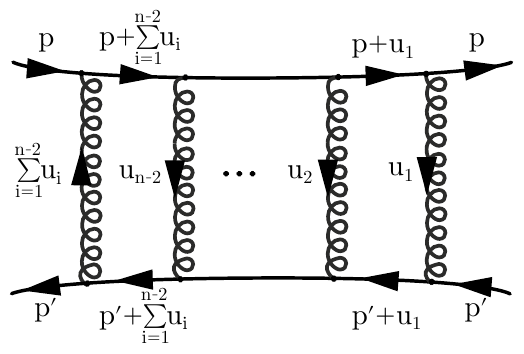} 
\end{center}
\caption{
   Example diagram for $n$-1 boson exchanges between two $n$-point quark loops with boson four-momenta $u_i$ and quark loop four-momenta $p$ and 
$p'$. Only a part of quark loops is shown here -- the full quark loops, photons connecting loops with incoming and outgoing asymptotic states, as well as the asymptotic 
states as shown in Figure~\ref{fig:fig1} are suppressed here for legibility.
}
\label{fig:fig2}
\end{figure}

To further explore the basic behavior of non-Abelian amplitudes in $q^2 \rightarrow 0$ limit, we step aside from Furry's theorem and attempt to calculate individual 
amplitude $Y_{\mu\nu}^f$.
After evaluating two quark loops 
and performing contraction due to bosonic fields, one is left with approximately three thousand terms where 
a scalar product $u_1\cdot u_2$ occurs inside PV 
coefficients. This makes it impossible to calculate the integral over $u_i$ exactly using tools for automatic calculations. 
To obtain at least a partial result, we may isolate terms 
free of scalar products and perform the integration over $u_i$ for these terms. This part of the result can be written in a form

  \begin{equation}
  \label{eq:Yg}
  Y_{\mu\nu}^f = g_{\mu\nu} \alpha_{\mathrm{s}}^{\frac{5}{2}}\bigg[ 
  \sum_{j=0}^{4} c_{0j} \log^j\frac{\mu^2}{\mb^2} + \frac{1}{\epsilon} \sum_{j=0}^{3} c_{1j} \log^j\frac{\mu^2}{\mb^2} + \frac{1}{\epsilon^2}\sum_{j=0}^{2} c_{2j} \log^j\frac{\mu^2}{\mb^2}
  + \mathcal{O}(\epsilon) \bigg],
  \end{equation}
where $\epsilon$ and $\mu^2$ regularize UV divergences and $c_{ij}=c_{ij}(m^2,\mb^2,\log(m^2/\mb^2),\mathrm{PV}(m,\mb))$ are rational functions in $m^2$ and $\mb^2$, 
polynomial in $\log(m^2/\mb^2)$ (up to the second order) and polynomial in PV functions (up to the second order). The maximal power of four of 
$\log({\mu^2}/{\mb^2})$ is connected with the presence of four off-shell four-momenta that are integrated over. The maximal power of two of $\log({m^2}/{\mb^2})$ is 
connected with integrating over two four-momenta of intermediate bosons with effective mass $\mb$. When 
performing the limit $\mb \ra 0$ on this rather complex result, one obtains zero. This \textit{infrared finite} behavior is not directly visible from complex 
coefficients $c_{ij}$ or from (\ref{eq:Yg}).
   In the following paragraphs, we will derive and generalize this infrared behavior for the analytical part of any two fermionic loop amplitude with
$n=n'$ and $n$ even in the $q^2 \rightarrow 0$ limit where $n-1$ bosons are exchanged between the two loops.  
  An example of such amplitude detailing the 
inter-quark-loop exchanges is shown in Figure~\ref{fig:fig2}{\color{blue}}. 
  For the abelian case, Furry's theorem leaves non-zero only amplitudes with the odd number of boson exchanges between the two quark loops ($n$ even, that is, even 
number of vertices with bosons when including also photon connecting the loop with external charge). For the non-abelian case, Furry's theorem cannot be 
applied due to the color part of the amplitude. Consequently, we are left also with amplitudes having an even number of boson exchanges between the two quark loops, i.e. 
$n$ odd. We perform the derivation only for $n$ even.

The Lorentz structure of one $n$-point fermionic loop calculation can be written as
  \begin{eqnarray}
  \nonumber
  A_{\mu_0,\mu_1 \cdots \mu_n} 
    &=& \int \mathrm{Tr}\bigg\{ \prod_{i=0}^{n-1} \gamma_{\mu_i} \frac{i}{\c{p} + \c{u}_i - m} \bigg\} \fd^4 p \sim \\
  \label{eq:p}
    &\sim& \prod_{j=0}^{n-1} \sum_{i=0}^{n-1} u_{i\mu_j} + \sum_{\sigma(\mu)} 
   \sum_{l=0}^{n/2-1} 
   \bigg[ 
                      \prod_{l'=1}^{2l+\delta}\bigg( \sum_{i=0}^{n-1}  u_{i\mu_{n-l'}} \bigg) 
                      \prod_{l'=1}^{(n-2l)/2} \bigg(  g_{\mu_{2l'-1}\mu_{2l'-2}} \bigg) 
   \bigg],
  \end{eqnarray}
where $u_n=0$, $u_0=0$ for the case of $q^2 \ra0$ limit, $\delta=0$ and $1$ for even and odd $n$, respectively, and sum with $\sigma(\mu)$ encodes all possible 
permutations of Lorentz indices $\mu_l$ (after expanding sums and products in (\ref{eq:p}), $\mu_0$ should be relabeled to $\mu$ to match the convention used in (\ref{eq:Y})). Equation~(\ref{eq:p}) says that after 
performing the integration over $p$, the result is a sum of all possible combinations of vectors $u_{i\mu_{l}}$ and metric tensors preserving the tensor structure of 
the original integral. Equation~(\ref{eq:p}), therefore, captures only the Lorentz structure of the solution of one loop integral. The full solution is given by not 
written special functions multiplying each term in the sum (\ref{eq:p}) (e.g. PV functions or hypergeometric functions \cite{Kershaw:1973km,Davydychev:1991va}). Some 
terms in the full solution may contain scalar products $u_i \cdot u_j$ as arguments of special functions. Some of these are non-analytical in $u_i\cdot u_j$, other 
are analytical, can be expanded in a series, and contribute the terms of the same form as in r.h.s. of (\ref{eq:p}). We will further discuss only the analytical part 
of the one-loop solution.

After contracting the solution of quark loops with numerators of boson propagators, we are left with the numerator, $\mathcal{N}$, of the 
multi-loop integral over $u_i$, which has the following form
\begin{equation}
\label{eq:numer1}
\sum_j \big[ g_{\mu\nu} c_1(u_i \cdot u_j, m) + u_{i\mu} u_{j\nu} c_2(u_i \cdot u_j, m) \big],
\end{equation}
where $j$ includes $i$ and $c_{1,2}$ are polynomials in scalar product $u_i \cdot u_j$ and polynomials in PV functions with arguments $u_i \cdot u_j, m$. 
When later integrating over a given four-momentum $u_i$, this numerator can be written as a sum of terms proportional to 
\begin{equation}
\label{eq:numer2}
  1, ~ u_{i\alpha}, ~ u_{i\alpha}u_{i\beta}, ~ u_{i\alpha}u_{i\beta}u_{i\gamma}, ... ,
\end{equation}
if excluding contributions non-analytical in $u_i \cdot u_j$ as discussed earlier.

In the case of an odd number of boson exchanges between two quark loops, there is inevitably at least one factor with the following integral over $u_i$,
\begin{equation}
\label{eq:denom1}
\int \frac{\mathcal{N}}
{
[(u_i+v)^2 - \mb^2][u_i^2 - \mb^2]
} \fd^4 u_i,
\end{equation}
where $v = \sum_{j,j \neq i} u_j$. This integration is then followed by at least one integral of the form
\begin{equation}
\label{eq:denom2}
\int \frac{\mathcal{N}}
{
u_j^2 - \mb^2
} \fd^4 u_j, ~ ~ \text{where } j \neq i.
\end{equation}
Performing the integral over $u_i$, implies integrating (\ref{eq:denom1}) along with numerators (\ref{eq:numer2}). 
The result may be written for a given numerator $\mathcal{N}$ as 
\begin{eqnarray}
\nonumber
\mathcal{N} \sim 1 &:& \sim 1 + \frac{1}{\epsilon} + \log\frac{\mu^2}{\mb^2}, \\
\nonumber
\mathcal{N} \sim u_{i\alpha} &:& \sim v_\alpha \bigg( 1 + \frac{1}{\epsilon} + \log\frac{\mu^2}{\mb^2} \bigg), \\
\nonumber
\mathcal{N} \sim u_{i\alpha} u_{i\beta} &:& \sim \bigg( v_\alpha v_\beta + \mb^2 g_{\alpha\beta} \bigg) \bigg( 1 + \frac{1}{\epsilon} + \log\frac{\mu^2}{\mb^2} \bigg), \\
\label{eq:uiint}
\mathcal{N} \sim u_{i\alpha} u_{i\beta} u_{i\gamma} &:& \sim \bigg( v_\alpha v_\beta v_\gamma + \sum_{\sigma(\alpha\beta\gamma)} \mb^2 v_\alpha g_{\beta\gamma} \bigg) \bigg( 1 + \frac{1}{\epsilon} + \log\frac{\mu^2}{\mb^2} \bigg), \\
\nonumber
&\cdots&.
\end{eqnarray}
  These are then integrated along with the denominator in (\ref{eq:denom2}) over each $u_j$. Terms in the numerator with odd powers of $u_j$ are equal to zero, and other terms lead to:
\begin{eqnarray}
\nonumber
\mathcal{N} \sim 1 &:& 
	\sim \mb^2 \bigg( 1 + \frac{1}{\epsilon} + \log\frac{\mu^2}{\mb^2} \bigg), \\
\nonumber
\mathcal{N} \sim u_{j\alpha} u_{j\beta} &:& 
	\sim \mb^4 g_{\alpha\beta} \bigg( 1 + \frac{1}{\epsilon} + \log\frac{\mu^2}{\mb^2} \bigg), \\
\label{eq:ujint}
\mathcal{N} \sim u_{j\alpha} u_{j\beta} u_{j\gamma} u_{j\delta} &:& 
	\sim \mb^6 \sum_{\sigma(\alpha\beta\gamma\delta)} g_{\alpha\beta} g_{\gamma\delta} \bigg( 1 + \frac{1}{\epsilon} + \log\frac{\mu^2}{\mb^2} \bigg), \\
\nonumber
&\cdots&,
\end{eqnarray}
  which may be generalized to 
\begin{equation}
\label{eq:ujintfull}
\mathcal{N} \sim \prod_{l=1}^{n} u_{j\mu_l} ~ ~ :  ~ ~\sim \mb^2 \sum_{l=1}^{n/2} \mb^{n-2l} \sum_{\sigma(\mu)} \prod_{l'=1}^{(n-2l)/2}g_{\mu_{2l'-1}\mu_{2l'}}
\bigg( 1 + \frac{1}{\epsilon} + \log\frac{\mu^2}{\mb^2} \bigg).
\end{equation}
The full numerator on the l.h.s. of (\ref{eq:ujint}) and (\ref{eq:ujintfull}) also includes $1/\epsilon$ and $\log(\mu^2/\mb^2)$ terms from (\ref{eq:uiint}). We will further concentrate only 
on the infrared part of the amplitude, which we label $f_M$. It can then be written as
\begin{equation}
  f_\mb(\alpha_s,\mb) \sim \alpha_s^{n-1} \big[ \mb^2( 1 + \log \mb^2 + \log^2 \mb^2) + \mathcal{O}(\mb^4) \big]. 
\end{equation}
Since $\lim_{\mb \ra 0} (\mb^2
\log^k(\mu^2/\mb^2)) = 0$ for any finite $k$, the infrared part of the amplitude is finite and non-zero for the non-abelian case.

Apparently, the above shown solution describes only a certain sub-class of amplitudes. As said before, we left out non-abelian amplitudes with $n$ odd. Further, intra-loop 
exchanges eliminate the applicability of Furry's theorem even for the abelian amplitudes and may leave various other contributions. The additional contributions may 
also come from amplitudes with self-interactions. Nevertheless, this particular solution allows us to write down a general form of the solution 
for a sum of all configurations with bosonic interactions between two quark loops as
\begin{eqnarray}
\label{eq:resabel}
Y_{\mu\nu} =   \bigg( g_{\mu\nu} - \frac{q_\mu q_\nu}{q^2} \bigg) \bigg(f_\mathbb{A}(\alpha,m,q) + f_\mathbb{N}(\alpha,m,q) \bigg) &...& \text{for abelian case,} \\
\label{eq:resnonabel}
Y_{\mu\nu} =   \bigg( g_{\mu\nu} - \frac{q_\mu q_\nu}{q^2} \bigg) \bigg(f_\mb(\alphas,\mb) + f_\mathbb{A}(\alphas,m,\mb,q) + f_\mathbb{N}(\alphas,m,\mb,q) \bigg) &...& \text{for non-abelian case,} 
\end{eqnarray}
where $f_\mathbb{A}$ represents the function analytical in $q$, $f_\mathbb{N}$ represents the non-analytical function (unsolved part of the amplitude), and $f_\mb$ 
is finite and non-zero if $\mb$ is non-zero. 
  One may see this as a trivial solution especially since no further information about $f_\mathbb{A}$ and $f_\mathbb{N}$ is provided, but importantly, this solution 
encodes the finding that the non-zero $M$ is a sufficient condition for a non-zero final amplitude in the non-relativistic limit (provided that $f_\mathbb{A} \neq - 
f_\mb$, which is discussed in the next section). This condition is not met in the abelian case. 
  As we only concentrate on the infrared part of the amplitude, the solution in (\ref{eq:resabel}) and (\ref{eq:resnonabel}) omits the short-distance ultraviolet part, 
which has a standard interpretation in terms of renormalizing bare quantities. While not providing much concrete information, the solution in (\ref{eq:resabel}) and 
(\ref{eq:resnonabel}) allows to resum the inter-quark-loop interactions, which is discussed in the next section. 

It is also important to discuss the value of $\alphas$ present in the amplitude. 
The $\alphas$ has a Landau pole at $q^2 = \Lambda_\mathrm{QCD}^2$ at the leading order. However, the realistic estimate of $\alphas$ for $q^2 \rightarrow 0$ is 
$\alphas(0) \approx 0.7-3$~\cite{Deur:2016tte}. 
Therefore, the value of $\alphas(0)$ should not lead to a change in the presented conclusions.


\section{Resumming inter-quark-loop interactions} 
\label{sec:res}

The result presented in the previous section represents the calculation of the interaction between only two quark loops. To obtain a full result, we need to resum at the 
operator level all the contributions of this type, which may take place anywhere in the photon. As we could see, these contributions contain finite terms in the 
non-abelian case. These contributions describe the strong interaction between hadronic degrees of freedom, which may lead to a physical force. This physical 
force based on the strong interaction is then different than the force based on the electromagnetic interaction. For this reason, we should not resum these 
contributions together with the free propagator of electromagnetic interaction, but we should resum them separately.

If we label free photon propagator times $(-a)$ as $1/X$ and $Y$ is defined in (\ref{eq:Y}), we perform the following resummation, 

\begin{equation}
	\label{eq:xyx}
\frac{1}{X} Y \frac{1}{X} +\frac{1}{X} Y \frac{1}{X} Y \frac{1}{X} + ... = \frac{1}{X} Y \bigg[ \frac{1}{X-Y}\bigg].
\end{equation}
Now, we can plug in the result at the eigenvalue level (\ref{eq:resnonabel}) and expand around $q^2 = 0$ to perform the $q^2 \rightarrow 0$ limit. When plugging 
(\ref{eq:resnonabel}) into (\ref{eq:xyx}) and expanding, one obtains
\begin{equation}
	\label{eq:M2}
i\mathcal{M}^{\alpha\beta} = - g^{\alpha\beta}\bigg( \frac{ia}{q^2} - \frac{1}{f_M(\alphas,\mb) + f_{\mathbb{A},0}(\alphas,m,\mb) + f_\mathbb{N}(\alphas,m,\mb,q)} + \mathcal{O}(q) \bigg), 
\end{equation}
where $f_{\mathbb{A},0}(\alphas,m,\mb)$ is the zeroth term in the expansion of $f_\mathbb{A}(q)$.
The $1/q^2$ behavior of the amplitude implies $1/r$ behavior of the resulting non-relativistic potential. It is noticeable that the non-analytical part, which is 
undetermined and which may be given, e.g., by a sum of polylogarithms, stands in the denominator along with $f_\mb$. If $\mb$ is non-zero, any behavior of 
$f_\mathbb{N}$ with $q^2 \rightarrow 0$ will keep unchanged the $1/q^2$ behavior of the amplitude.  Namely, for $f_\mathbb{N} \rightarrow \infty$ in 
the limit of $q^2 \rightarrow 0$, all the terms but the first one vanish. For $f_\mathbb{N}$ finite or zero in the limit of $q^2 \rightarrow 0$, only the first 
term in (\ref{eq:M2}) is divergent, and the second finite term represents a finite contribution leading to a local correction of the $1/r$ behavior of the potential.  Terms with 
$1/\sqrt{q^2}$ or $\ln q^2$, which are divergent for $q^2 \rightarrow 0$ and which would give rise to corrections of potential behaving as $1/r^2$ or $1/r^3$, 
respectively \cite{Helayel-Neto:1999ryv}, are absent.

We should note that $f_\mathbb{N}$ is present in the numerator not only with $f_\mb$ but also with $f_{\mathbb{A},0}$.
We did not prove in general that $f_{\mathbb{A},0} \neq -f_\mb$, which would lead to a cancellation and possible deviation from $1/r$ behavior due 
to the presence of $1/f_\mathbb{N}(q)$ in the amplitude, but we verified that $f_{\mathbb{A},0} \neq -f_\mb$ for amplitudes discussed in Section~\ref{sec:asy}.

There are two other observations that can be made. First, one can compare the amplitude (\ref{eq:M2}) to the leading order scattering amplitude for the photon 
exchange between two fermions of the same electric charge, which reads $i\mathcal{M}^{\alpha\beta} = g^{\alpha\beta}\frac{ia}{q^2}$ (still ignoring the description 
of the observable initial and final state which will be discussed in the next section). From this comparison, one can see that there is a change in the sign 
between the two amplitudes which originates in the resummation (\ref{eq:xyx}) and subsequent expansion in $q$. The details about the sign and strength of the interaction 
are then discussed in the next section. Second, one can easily verify that the sign of $Y$ influences only the sign of the second and other terms in (\ref{eq:M2}). 
The sign of contributions to $Y$ is indeed alternating due to the presence of different number of interacting fields, due to the color algebra, or due to the presence of 
amplitudes with crossings.


\section{Sign and strength of the interaction}   
\label{sec:sign}

We will now turn attention to the problem of the sign of the interaction. First, we will briefly recap the origin of the sign in two basic theories -- in Yukawa and 
electromagnetic theory. As this should be well known from classical textbooks (see e.g.~\cite{Peskin:1995ev}), we will recap that in the form of a table where 
the multiplicative sign factors and their source are shown for interaction between two fermions. The origin of sign of interaction for fermion-fermion (\papa{f}{f}), 
fermion-anti-fermion (\papabar{f}{f}), and two different fermions (\papa{f_{1}^{-}}{f_{2}^{+}} and \papa{f_{1}^{-}}{\bar{f}_{2}^{-}}) is the following:
\begin{center}
	\begin{tabular}{   c|c|c||c|c|c|c|  } \cline{2-7}
		      & \multicolumn{2}{c||}{Yukawa} & \multicolumn{4}{c|}{electromagnetic} \\ \cline{2-7} 
		      & \papashift{f}{f}  & \papabarshift{f}{f}  &  \papashift{f}{f}  & \papabarshift{f}{f}  &  \papashift{f_{1}^{-}}{f_{2}^{+}}  & 
                                                                                                                \papashift{f_{1}^{-}}{\bar{f}_{2}^{-}} \\ \hline \hline 
		 \multicolumn{1}{|c|}{\textit{fermion current}}       & $+1$ & $-1$ & $+1$ & $+1$ & $+1$ & $+1$ \\ \hline  
		 \multicolumn{1}{|c|}{\textit{intermediate particle}} & $+1$ & $+1$ & $-1$ & $-1$ & $-1$ & $-1$ \\ \hline 
		 \multicolumn{1}{|c|}{\textit{normal product}}        & $-1$ & $+1$ & $-1$ & $+1$ & $-1$ & $-1$ \\ \hline 
                 \multicolumn{1}{|c|}{\textit{charge}}                & $+1$ & $+1$ & $+1$ & $+1$ & $(-1)(+1)$ & $ (-1)(-1)$ \\ \hline \hline
                 \multicolumn{1}{|c|}{$(-1) ~ \cdot ~$ total}      & $+1$ & $+1$ & $-1$ & $+1$ & $+1$ & $-1$\\ \hline \hline
	\end{tabular}
\end{center}

The \textit{fermion current} sign comes from the normalization of bispinors in $\bar{\Psi}\Psi$ and $\bar{\Psi}\gamma^\mu\Psi$ currents in Yukawa and 
electromagnetic interaction, respectively. In the non-relativistic limit, only the term with $\gamma^0$ remains due to the definition of bispinors, which compensates the 
sign from the normalization of bispinor in the case of electromagnetic interaction between fermion-antifermion. Since only $\gamma^0$ term is active in fermion currents, 
only $g_{00}$ term remains in the photon propagator, leading to the repulsive instead of attractive interaction in the case of electromagnetic interaction between 
\papa{f}{f} pair (\textit{intermediate particle} sign). While the minus sign from the \textit{normal product} of creation and annihilation operators compensates the 
sign from bispinor normalization in the Yukawa case, it leads to the total attractive force in the case of \papabar{f}{f} electromagnetic interaction. Since $f$ and 
$\bar{f}$ represent two states of one wave function, the sign of the \textit{charge} in the coupling remains the same and it is the order of creation and 
annihilation operators in the \textit{normal product} which dictates the alternating sign of the corresponding Coulomb potential. In the case of 
\papa{f_{1}^{-}}{f_{2}^{+}}, the order of creation and annihilation operators is the same as in the \papa{f}{f} case. The order is the same even in the case of 
exchanging $f_2$ by its antiparticle since $\bar{f_2}$ is not an antiparticle to $f_1$. In the case of \papa{f_{1}^{-}}{f_{2}^{+}} and 
\papa{f_{1}^{-}}{\bar{f}_{2}^{-}}, the total sign of electromagnetic interaction is dictated by the sign of \textit{charge} in the coupling. The factor $(-1)$ in 
front of the total sign comes from $q^2$ being space-like, but in general, it is only the relative sign that can be determined while the total sign is a matter of 
convention.

In the case of residual strong interaction, the signs due to \textit{fermion current} and \textit{normal product} do not change with respect to the electromagnetic 
interaction. Neither changes the sign due to the presence of $g_{00}$ in the non-relativistic limit. What changes is the overall sign, which is due to resummation as 
discussed in Section~\ref{sec:res}. Since the physical interaction is due to the strong interaction which does not distinguish charge and which probes the strongly 
interacting degrees of freedom in the wave function of incoming fermions, the sign of the electromagnetic charge is not relevant. The sign in coupling is thus the 
same as in the Yukawa case leading to the attractive interaction for all the configurations except for the case of \papabar{f}{f} pair where it is repulsive. 
This repulsive behavior in \papabar{f}{f} case has the same origin as the attractive behavior in the electromagnetic interaction of \papabar{f}{f}, which is due to the 
description of physical, asymptotic states as discussed in previous paragraphs. This prediction represents a distinguishable behavior of the residual force.

Since our interaction probes only hadronic degrees of freedom in the fermion's wave function, the magnitude of coupling is given by the hadronic part of 
electromagnetic coupling. 
The interaction strength of strongly 
interacting part of electromagnetic fields can be determined from the leading order expansion of the electromagnetic coupling constant $\alpha(q^2)$ which reads
\begin{eqnarray} 
	\label{eq:est1}
	\alpha(q^2) = \frac{\alpha_0}{1 - \Delta\alpha(q^2)} = \alpha_0 [ 1 + \Delta\alpha(q^2)_\mathrm{had} + \\ \nonumber
	+ \Delta\alpha(q^2)_\mathrm{lep} + \Delta\alpha(q^2)_\mathrm{top} + \mathcal{O}(\Delta\alpha^2)]. 
\end{eqnarray} 
This gives an estimate of the strength of interaction mediated by a photon containing quark loops, which is
\begin{equation} 
	\label{eq:est2}
	\alpha_0 \Delta\alpha(q^2)_\mathrm{had} \equiv a(q^2).
\end{equation} 
  The $a(q^2)$ is then the value of coupling to be used in resummed expressions (\ref{eq:M2}).


\section{Estimates of the magnitude of residual force at macroscopic scales}
\label{sec:est}

As discussed in previous sections, the residual force may be expected to act as an imperceptible, attractive component of the Coulomb potential, which does not 
distinguish the sign of the charged object. It may therefore be detectable at macroscopic scales as a difference in the magnitude of the repulsive Coulomb force 
between two charged objects with equal charges $e_1=e_2=e$ and the magnitude of the attractive Coulomb force between two opposite charges with $e_1=-e_2=e$.

We estimate the ratio of magnitudes of the residual force and electromagnetic force, $\alphares/\alpha_0$, using the values of $\alphah$ based on numerical 
calculations from~\cite{alphaQED}, which use extrapolations from hadron measurements as briefly discussed in Section~\ref{sec:sign} and detailed in 
Ref.~\cite{Jegerlehner:2019lxt}. The result is $10^{-20}$ and $10^{-26}$ for the eV scale and meV scale, respectively. That is, for these scales, the ratio is not 
extremely far from the ratio of the magnitude of gravity and electromagnetic force for two protons (which is $10^{-36}$). We should stress that the presented numbers 
provide only a basic estimate based on extrapolations of trends in the evolution of $\alphah$ done over several orders of magnitude, and as such, they should be taken with 
a reserve. Precise estimates of $\alphah$ based on lattice QCD would be needed to build a firm quantitative statement. Nevertheless, we can compare these estimates 
with the precision of current terrestrial experiments.

Assuming that terrestrial experiments with macroscopic electromagnetic forces can be done around the energy level of eV, detecting the residual force at 
terrestrial conditions would require precision of the measurement of electromagnetic processes at the level of $10^{-20}$. The relative uncertainty of the knowledge of 
elementary charge and permittivity of vacuum is at the level of $10^{-10}$ \cite{codata2018}. The precision of the measurement of Coulomb force for elementary charges is at 
a similar level \cite{Peng2020}. Thus, it seems impossible at the present time to verify or falsify the presence of residual force by terrestrial experiments.

While it is not possible to measure the residual force at terrestrial scales, it may play a role in large-scale astrophysical objects due to the presence of QED 
plasma in thermal equilibrium. Two examples of those objects are the warm ionized medium (WIM) in interstellar space 
\cite{Haffner:2009mh,Cox2005,Vogelsberger:2019ynw} and intracluster medium (ICM) in clusters of galaxies \cite{Werner2020,Simionescu2019}.
  Those media are composed predominantly of ionized hydrogen. They are
macroscopically electrically neutral due to Coulombic interactions between ions and electrons acting 
within a Debye volume defined by the Debye length, $\lambda_\mathrm{D}$. 
The screened electrostatic potential is given by \cite{Bittencourt}
\begin{equation}
\Phi_\mathrm{deb}(r) = \frac{1}{4\pi\epsilon_0} \frac{e}{r} \exp \bigg( -\frac{\sqrt{2}r}{\lambda_\mathrm{D}} \bigg),
\label{eq:Phi}
\end{equation}
where $e$ is the unit charge.
While the QED interactions among positive and negative charges lead to net-zero Coulomb force among plasma constituents, the residual force is attractive for both the 
positive and negative charges. 
The net-zero Coulomb force may, therefore, allow the residual force to be macroscopically distinguishable. 
The estimate of the magnitude of the residual force between a given H$^+$ ion and other ions in the Debye volume can be obtained by replacing 
$\alpha_0$ with $\alphares(q^2)$ in a classical calculation of force due to potential (\ref{eq:Phi}) which is 
\begin{eqnarray}
\nonumber
F_\mathrm{res}(\lambda_\mathrm{D}) = 2 n_i e \int^{\lambda_\mathrm{D}} \frac{\partial \Phi_\mathrm{deb}(r)}{\partial r} \mathrm{d}^3 x \bigg|_{\alpha_0 \rightarrow \alphares(q^2)} = \\ 
= 4 \pi [\sqrt{2} - (1 \! + \! \sqrt{2})\exp(-\sqrt{2})] n_i \alphares(q^2) \lambda_\mathrm{D} \hbar c, ~ ~
\label{eq:Fres}
\end{eqnarray}
where $n_i$ is the number density of the ionized medium. 
The magnitude of residual force may be compared with the magnitude of gravitational force within the same volume
\begin{equation}
F_\mathrm{G}(\lambda_\mathrm{D}) = 4 \pi n_i \frac{m^2_{\mathrm{H}^+}}{m_\mathrm{Pl}^2} \lambda_\mathrm{D} \hbar c ,
\end{equation}
where $m_\mathrm{Pl}$ is the Planck mass and $m_{\mathrm{H}^+}$ is the proton mass. 

For WIM, where $n_i \sim 10^{-1}$~cm$^{-3}$, $\lambda_\mathrm{D} \sim 10$~m, and temperature $T \sim 10^{4}$~K ($q^2$ at the scale of eV$^{2}$), the magnitude 
of $F_\mathrm{res}$ is $10^{-46}$~GeV/fm. 
For ICM, where $n_i \sim 10^{-3}$~cm$^{-3}$, $\lambda_\mathrm{D} \sim 10^{3}$~m, and $T \sim 10^7 - 10^8$~K ($q^2$ at the scale of keV$^{2}$), the order of magnitude of 
$F_\mathrm{res}$ is $10^{-40}$~GeV/fm. The order of magnitude of $F_\mathrm{G}$ is $10^{-62}$ and $10^{-61}$~GeV/fm for WIM and ICM, respectively.
  While the values of $F_\mathrm{res}$ should be taken with reserve as they are based on extrapolations of $\alphah$, as previously discussed, they suggest 
that the magnitude of $F_\mathrm{G}$ might be significantly smaller, implying that the presence of attractive force may be stronger than the self-gravity within the 
Debye volume. This might then have consequences on the understanding of the dynamics of these media.\footnote{
  In fact, one may speculate that the residual force may be somewhat more universal since all charged particles, even those bound in multipoles, may be emitting 
  soft photons of large-wave lengths ($\lambda$), which do not interact with charges in other multipoles (having size $r$, $r \ll \lambda$) as they cannot recognize 
  the structure of the multipole. However, these photons may still be carriers of the residual force. Further, all particles contain electromagnetic degrees of 
  freedom in the wave function and may therefore be sensitive to this force (see e.g. review in Ref.~\cite{SajjadAthar:2022pjt} for the case of neutrinos).
}

\section{Discussion and conclusion}
\label{sec:discussion}

Residual forces of electromagnetism and strong interaction give rise to many structures in nature, but their description by a theory from first principles is not
simple. Here, we proposed a new residual force, which was not discussed previously in the literature, the residual QCD force in electromagnetically interacting fields.  
We calculated the asymptotic behavior of relevant scattering amplitudes, performed their resummation, and analyzed the sign 
of the resulting interaction. The non-relativistic potential of the residual force exhibits $1/r$ behavior and it is attractive for all particles except for the interaction between fermion and anti-fermion. Then, we calculated the primary experimentally observable consequences of this Standard Model force. 
While imperceptible at the scale of terrestrial experiments, the presence of this force may have an impact at astrophysical scale, in particular, it may influence the dynamics of  
the warm ionized medium present in galaxies and the intergalactic space.

Better quantitative understanding of the behavior of hadronic vacuum polarization, e.g. based on lattice QCD calculations, may bring further insight. Including an 
attractive force of this kind into the simulation of large-scale structures in the universe that involve QED plasma may also be fruitful. The presented paper opens 
the door for these investigations.

\section*{Acknowledgment}

I would like to thank to Ji\v r\' i Ho\v rej\v s\' i for reading the manuscript and for many useful discussions.
Further, I would like to thank to  
Ji\v r\' i Dolej\v s\' i, David Heyrovsk\' y, Ond\v rej Pejcha,
Ji\v r\' i Novotn\' y, Michal Malinsk\' y, Karol Kampf, Fred Jegerlehner, and Alfredo Iorio
for useful discussions on this subject or for technical help.
This work was supported by Grant Agency of the Czech Republic under grant 
22-11846S and by project ERC-CZ LL2327. 

\bibliography{paper}
\bibliographystyle{elsarticle-num}
\end{document}